\documentclass[12pt,preprint]{aastex}

\newcommand{\myemail}{adamwr@uiuc.edu}
\slugcomment{submitted to \apj  \ 31 Oct 2003, accepted 24 Jan 2004}
\shorttitle{Discovery of Variable Quasars}
\shortauthors{Rengstorf, A.W. et al.}

\begin{document}

\title{New Quasars Detected via Variability in the QUEST1 Survey}

\author{A.W. Rengstorf\altaffilmark{1,2,8}, S.L. Mufson\altaffilmark{2}, C. Abad\altaffilmark{6}, B. Adams\altaffilmark{2}, P. Andrews\altaffilmark{5}, C. Bailyn\altaffilmark{4}, C. Baltay\altaffilmark{4,5}, A. Bongiovanni\altaffilmark{6}, C. Brice\~{n}o\altaffilmark{6}, G. Bruzual\altaffilmark{6}, P. Coppi\altaffilmark{4,5}, F. Della Prugna\altaffilmark{6}, W. Emmet\altaffilmark{5}, I. Ferr\'{i}n\altaffilmark{7}, F. Fuenmayor\altaffilmark{7}, M. Gebhard\altaffilmark{3}, J. Hern\'{a}ndez\altaffilmark{7}, R.K. Honeycutt\altaffilmark{2}, G. Magris\altaffilmark{6}, J. Musser\altaffilmark{3}, O. Naranjo\altaffilmark{7}, A. Oemler\altaffilmark{4,12}, P. Rosenzweig\altaffilmark{7}, C.N. Sabbey\altaffilmark{4,11}, Ge. S\'{a}nchez\altaffilmark{6}, Gu. S\'{a}nchez\altaffilmark{6}, B. Schaefer\altaffilmark{5,9}, H. Schenner\altaffilmark{6}, J. Sinnott\altaffilmark{5,10}, J.A. Snyder\altaffilmark{4,5}, S. Sofia\altaffilmark{4}, J. Stock\altaffilmark{6}. W. van Altena\altaffilmark{4}, and A.K. Vivas\altaffilmark{6}}

\altaffiltext{1}{current address: Dept. of Astronomy, University of Illinois, 1002 W. Green St., Urbana, IL 61801, \myemail}
\altaffiltext{2}{Dept. of Astronomy, Indiana University, 727 E. 3rd St., Bloomington, IN 47405}
\altaffiltext{3}{Dept. of Physics, Indiana University, 727 E. 3rd St., Bloomington, IN 47405}
\altaffiltext{4}{Dept. of Astronomy, Yale University, P.O.Box 208101, New Haven, CT 06250-8101}
\altaffiltext{5}{Dept. of Physics, Yale University, P.O.Box 208121, New Haven, CT 06250-8121}
\altaffiltext{6}{Centro de Investigaciones de Astronom\'{i}a, Apartado Postal 264, M\'{e}rida 5101-A, Venezuela}
\altaffiltext{7}{Universidad de Los Andes, Apartado Postal 26, M\'{e}rida 5251, Venezuela}
\altaffiltext{8}{Visiting Astronomer, Kitt Peak National Observatory, National Optical Astronomy Observatory, which is operated by the Association of Universities  for Research in Astronomy, Inc. (AURA) under cooperative agreement with the  National Science Foundation.} 
\altaffiltext{9}{current address: Department of Physics and Astronomy, Louisiana State University, Baton Rouge, LA 70803-0001}
\altaffiltext{10}{current address: Cornell University, Ithaca, NY 14853-1501}
\altaffiltext{11}{current address: Bogle Investment Management, Wellesley, MA 02481}
\altaffiltext{12}{Carnegie Observatories, 813 Santa Barbara Street, Pasadena, CA 91101-1292}

\begin{abstract}
By observing the high galactic latitude equatorial sky in drift scan mode with the QUEST (QUasar Equatorial Survey Team) Phase 1 camera, multi-bandpass photometry on a large strip of sky, resolved over a large range of time scales (from hourly to biennially) has been collected. A robust method of ensemble photometry revealed those objects within the scan region that fluctuate in brightness at a statistically significant level. Subsequent spectroscopic observations of a subset of those varying objects easily discriminated the quasars from stars. For a 13-month time scale, $38\%$ of the previously known quasars within the scan region were seen to vary in brightness and subsequent spectroscopic observation revealed that approximately $7\%$ of all variable objects in the scan region are quasars. Increasing the time baseline to 26 months increased the percentage of previously known quasars which vary to $61\%$ and confirmed via spectroscopy that $7\%$ of the variable objects in the region are quasars. This reinforces previously published trends and encourages additional and ongoing synoptic searches for new quasars and their subsequent analysis. During two spectroscopic observing campaigns, a total of 30 quasars were confirmed, 11 of which are new discoveries and 19 of which were determined to be previously known. Using the previously cataloged quasars as a benchmark, we have found it possible to better optimize future variability surveys. This paper reports on the subset of variable objects which were spectroscopically confirmed as quasars.
\end{abstract}


\section{Introduction}
After the first all-sky radio surveys were conducted in the late 1950s, \citet{Matthews63} found the first quasar in a point-like optical counterpart for a 3C radio object and also noted that several 3C objects exhibited continuum brightness fluctuations. However, it was not until ten years later that it was suggested that quasars might be detected solely by means of this intrinsic variability \citep{Vandenbergh73}. Another decade passed before the first sample of quasars discovered solely by their variability was assembled \citep{Hawkins83}. 

Over the past two decades, more than a dozen major surveys have been carried out on the variability of optical- or radio-selected samples of quasars \citep[][and references therein]{Helfand01}. Depending on the exact selection criteria used and the photometric accuracy and limiting magnitude of the survey, results range from 50\% of quasars exhibiting variability of at least 0.15 magnitudes on a two-year time scale \citep{Cimatti93} to practically all quasars exhibiting variability of at least 0.05 magnitudes on a 15-year time scale \citep{Trevese94,Veron95}. Other studies, using proper motion in addition to variability as selection criteria \citep[eg][]{Meusinger02}, find that $80\%$ of their sample of quasars show detectable variability with approximately $0.1$ mag photometric accuracy over a baseline of 14 years \citep{Kron81} and all quasars exhibit variability over a 16-year baseline \citep{Majewski91}. 

The major difference between QUEST and these previous quasar variability studies is that we are using the variability of quasars as measured by QUEST as a means of discovery rather than using a known sample of quasars to study their variability characteristics. In addition to QUEST, there exist other current large-scale surveys which are making significant advancements in the discovery and study of quasars, both through means of variability and through chromatic selection processes. Recently, MACHO \citep{Geha03} and OGLE \citep{Dobrzycki03} have found quasars behind the Magellanic Clouds by means of their intrinsic variability. The Sloan Digital Sky Survey (SDSS) \citep[eg.,][]{Schneider03,VandenBerk03,deVries03} has also made great strides in increasing the number of known quasars. The time domain, however, has not to date been sampled as robustly by SDSS as by QUEST. SDSS has at most a handful of observations of any given area on
the sky. In contrast, QUEST has sampled a smaller area (roughly $200\ deg^2$) on the high galactic latitude sky as frequently as two times per night over the course of three observing seasons, capturing 69 images of the same region. It is with the QUEST repeat observations and more robust light curves that we are able to investigate a realm of phase space distinct from other quasar surveys. 

This work investigates the notion that \emph{all} quasars, by nature of their presumed central engine, vary in brightness at some level on a time-scale of a few to several years \citep[see][for a review of AGN central engine variability]{Ulrich97}. As more and more all-sky surveys are undertaken and as cosmological tests use quasars as a means to compute cosmological parameters, it becomes paramount to understand the selection effects and biases of existent quasar catalogs. By comparing quasar catalogs compiled via optical colors or radio flux to a sample of quasars collected from achromatic processes such as variability or proper motion, we can better ensure that we have a homogenous sample of quasars with understood biases. This paper reports on the quasars that were found by QUEST via photometric variability and spectroscopic confirmation. The details of the data reduction and analysis and the variability criteria are discussed in a subsequent paper \citep{Rengstorf04}.
 
In \S 2, we discuss the photometric and spectroscopic data collection. \S 3 discusses the variable quasars in the survey region, both those confirmed by this work and those previously cataloged. \S 4 discusses the findings and introduces some ideas for future work in this area and \S 5 presents our conclusions.

\section{Observations}

The QUEST Phase 1 camera \citep{Baltay02} is installed on the 1m-Schmidt telescope at the Llano del Hato Observatory\footnote[1]{The Llano del Hato observatory is operated by the Centro de Investigaciones de Astronom\'{i}a for the Ministerio de Ciencia y Tecnologia of Venezuela.}, Venezuela. The
camera consists of 16 (a 4x4 array) front-illuminated 2048 x 2048 15
\micron\ pixel CCDs. The camera is situated at the prime focus of the
Schmidt telescope. With a focal length of 3 meters, the plate scale is approximately 1\farcs03 per pixel. Due to gaps between
each row of four chips, the camera has a total field of view of
2\fdg4\ x\ 3\fdg5 and a total sky coverage of 
$5.53\ deg^2$.

The telescope is normally operated in drift scan mode, remaining
parked at a fixed hour angle (HA), but not necessarily HA = 0. The CCD
chips are then clocked out at the sidereal rate to prevent images from
blurring across the chip. This observing procedure allows for an integration time of 143 seconds
for each chip along the celestial equator and results in limiting
magnitudes of $B = 18.5$, $V = 19.2$, and $R = 19.5$ at a
signal-to-noise ratio of 10 \citep{Baltay02}. Each row of four chips has a broadband filter placed over it. Every object passing across the chip array can therefore be imaged through
up to four different broadband filters. The variability scans used a set of R,B,R,V \citep[refer to][for detailed reviews of QUEST and the Phase 1 camera]{Snyder98,Baltay02}.

\subsection{Photometric Scans}

To date, the QUEST variability survey has 69 usable scans of varying quality. These scans
cover a strip of sky approximately 2\fdg4 wide in declination (centered at -1\degr) and $5.5$ hours wide in right ascension at high
galactic latitude ($10^h <$ RA $< 15^h30^m$). All scans were taken over the course of three observing seasons at the Llano del Hato Observatory. (1999 February - March, 2000 February - March and 2001 March - April).

Using a suite of custom software, data reduction, object detection, astrometry and aperture photometry were performed and a catalog for each night was created. A series of cuts and tests were made on objects from each individual night. Objects were matched together among all the scans and a second series of tests and cuts were performed on the ensemble of objects and exposures. An ensemble photometric solution was used to determine a list of variability candidates based on the confidence level (CL), i.e., the percentage chance that an object is not variable due to random fluctuation. This algorithm is performed on each chip individually.

After passing all of the above tests and cuts, the results from the individual chip solutions are used to compile a final list of variable objects. Whether an object is ultimately determined to be variable
depends both on the number of chips on which it was detected and the subset of those chips on which it was seen to vary. As a self-consistency check, any object detected on two or more chips is required to vary on \emph{at least} the two R chips to be included in the variable object list. Any object only detected on one chip is disregarded, regardless of its variability characteristics.

Two lists of variables were compiled for each of the two time baselines:
one list of objects which varied as described above at the 95\% CL and a list of objects which varied at the 85\%
CL. With 13 months of data, a total of 301,472 ensemble
objects were detected. Out of those objects, 2,631 were variable at the 85\% CL and
1,580 were variable at the 95\% CL. With the addition of another epoch
of data, the baseline expanded to 26 months and 280,538
ensemble objects were detected, 2,051 of which were accepted at the 85\% CL and
1,356 of which were accepted as variables at the 95\% CL. While leaving the details of the data reduction algorithms and candidate selection process to be presented in \citet{Rengstorf04}, it is worth noting that the slight difference in the number of detected objects in the 13-month and 26-month data sets is due to a cut in the data which required an object to appear on a minimum percentage of the valid exposures. 

\begin{figure}
\plotone{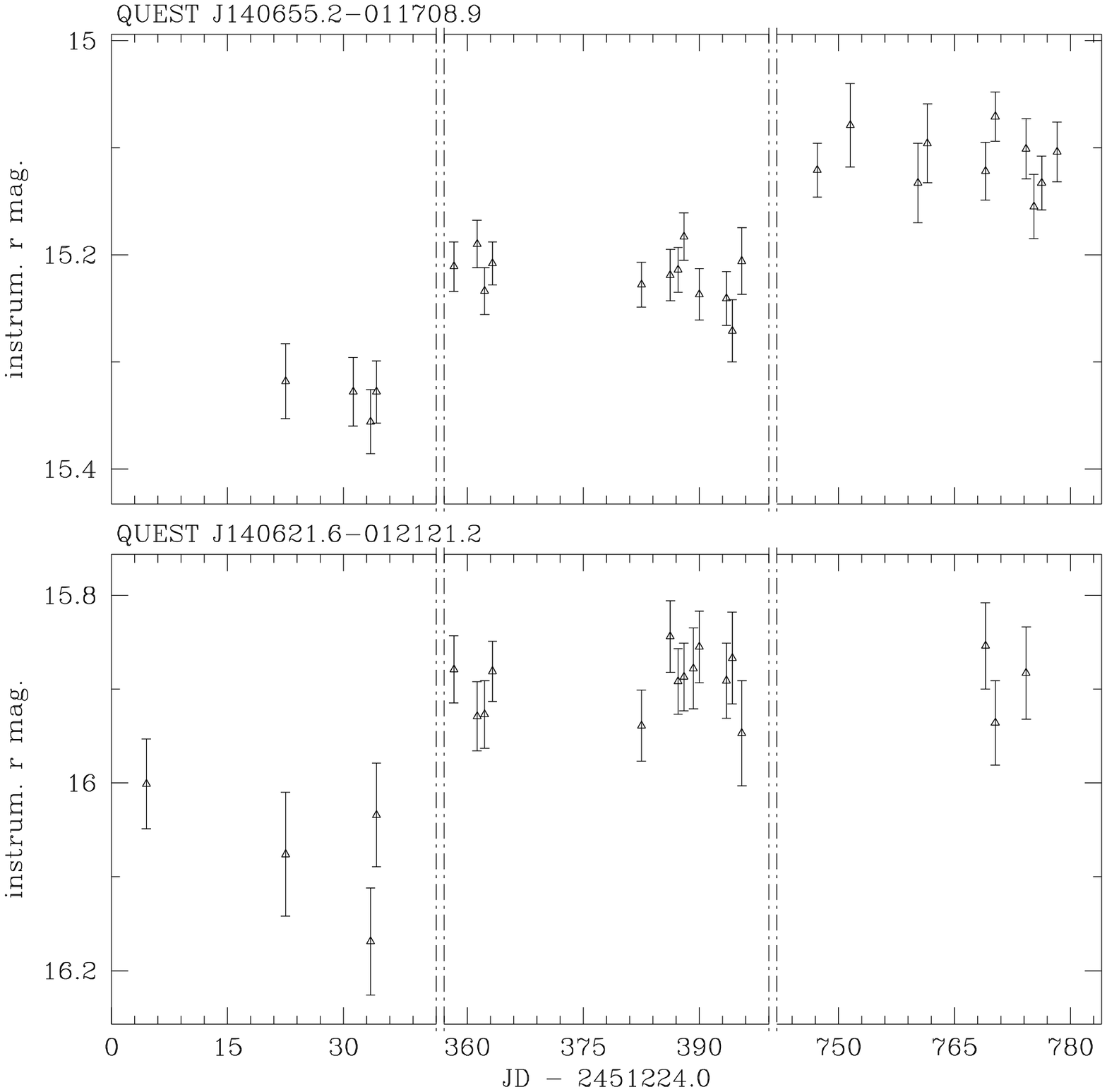}
\caption[Sample light curves for variable quasars]{The light curves (instrumental r magnitude) of two representative variable objects later confirmed via spectroscopy to be quasars (QUEST J140655.2-011708.9 and J140621.6-012121.2). The three sub-panels in each plot represent 45-day spans in Feb.-Mar. 1999, Feb.-Mar. 2000 and Mar.-Apr. 2001.}
\label{lc}
\end{figure}

Lightcurves for two representative variability candidates which were later confirmed as quasars are given in Fig.\,\ref{lc}. Note the characteristic variability signature, which in both cases exhibits low-level fluctuations on short (i.e., monthly) time scale and larger fluctuations on an annual time scale.

\subsection{Spectroscopic Observations}

Once all the variable objects were identified, the next step was to
separate the quasars from stars and other variable objects. While
stellar in appearance, quasars generally have very distinctive broad emission
lines (BEL) which easily distinguish them from variable stars. In
order to identify which of the objects in the candidate list are
quasars, spectra for each of the candidates are required. This was
achieved by using the Hydra multi-object spectrograph (MOS) on the
3.5-meter WIYN\footnote[2]{The WIYN Observatory is a joint facility of the University of Wisconsin-Madison, Indiana University, Yale University, and the National Optical Astronomy Observatory.} telescope at Kitt Peak National Observatory. The MOS at WIYN is able to
obtain spectra for several dozen objects at once, making it an ideal
instrument for identifying large numbers of variable objects. Those
objects which are both time-variable and have detectable BEL are
considered to be quasars.

Initial observations were obtained over four nights between 2001 April 17 and April 21. The WIYN 316@7.0 grating (316 lines per millimeter with a blaze angle of
$7\degr$ and a plate scale of approximately $0.23$ nm per pixel)
was used to obtain low resolution spectroscopy over a wide range of
wavelengths ($440$ nm $< \lambda < 915$ nm). The GG-420 filter was
used to filter out light shortward of $424$ nm, avoiding overlap with the
second-order spectra in the blue. The spectra are limited on the red
end by both the decreased quantum efficiency of the detector beyond
$1000$ nm and the large amount of atmospheric contamination in the
near infrared. 

The Hydra fields were selected from within the photometric scan region based on spatial clumping of variable objects so as to maximize the number of spectra obtained per field. Nine Hydra fields were imaged for a total of one hour each ($3$
exposures $\times\ 20$ min. each). These 9 fields cover approximately
$7.1\ deg^2$ of the sky and contain a total of 248 candidate
objects. Of these 248 candidates, 17 ($6.9\%$) were determined to be quasars. Additional observations were obtained over three nights between 2002 February 14 and February 16. The Hydra configuration described above for the 2001 observing
run is identical to the configuration used during the 2002 observing
campaign. Twelve Hydra fields were imaged for a total of one hour each
($3$ exposures $\times\ 20$ min. each). These 12 fields cover
approximately $9.4\ deg^2$ of the sky and contain 203 candidate
objects. Of these 203 candidates, 15 ($7.4\%$) were determined to be quasars. Table\,\ref{fields} reports the details for each of the fields observed during both observing runs.
Each spectrum is classified as one of the following: a quasar (Q), showing distinctive broadened emission lines, a stellar spectrum (S), showing a characteristic blackbody-like curve over the observed wavelength range, an M dwarf star (M), showing the broad molecular absorption bands characteristic of cool M stars, and spectra with low or no noticeable signal or those that do not otherwise fit into the three categories (?) described above.

\begin{figure}
\plotone{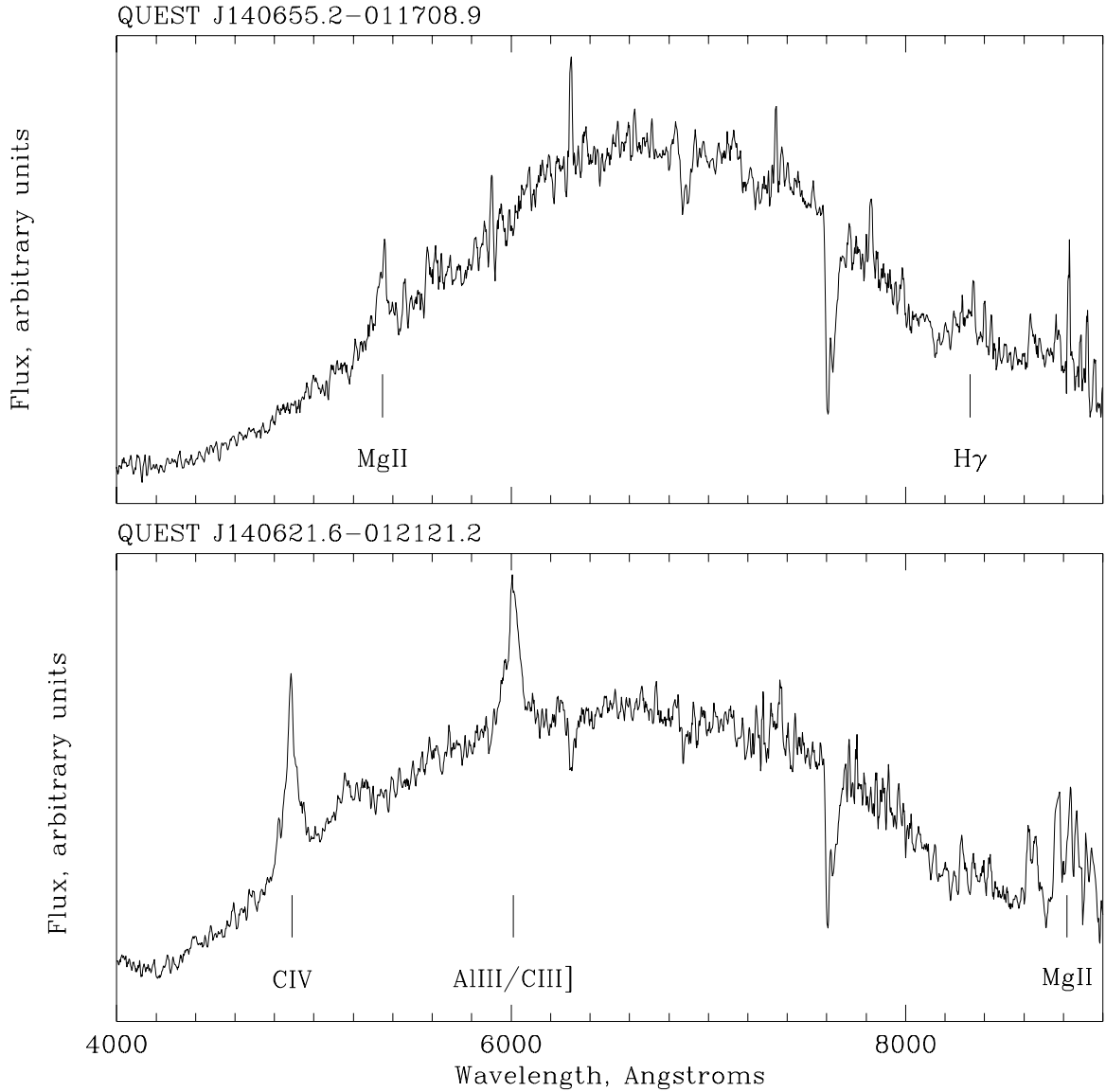}
\caption[Sample spectra for variable quasars]{Low resolution Hydra spectra of two QUEST variable quasars. J140655.2-011708.9 is reported as a new quasar ($z=0.91$). J140621.6-012121.2 ($z=2.15$) has been cross-identified with a 2dF QSO Redshift Survey quasar.}
\label{spectra}
\end{figure}

The spectroscopic data were reduced and analyzed using the \emph{dohydra} IRAF\footnote[3]{IRAF is distributed by the National Optical Astronomy Observatories, which are operated by the Association of Universities for Research in Astronomy, Inc., under cooperative agreement with the National Science Foundation.} package. Two representative spectra, corresponding to the light curves shown in Fig.\,\ref{lc}, are presented in Fig.\,\ref{spectra}. Both show the prominent broadened emission lines used to identify the QUEST variable quasars.

\section{Variable Quasars}

It was possible to study variable quasars in two distinct ways: by spectroscopically confirming a subset of the candidates chosen via photometric variability and by comparing the list of candidates to a catalog of known quasars. The results of both methods are presented below.

\subsection{Variable Quasar Discovery}

The analysis of the reduced Hydra spectra was straightforward. In essence, it involved the visual inspection of the
individual spectra. A quasar will make itself known with the presence
of BEL features, generally several hundred Angstroms in equivalent
width, on top of a roughly power-law continuum emission, which may or
may not be in evidence. (A strong continuum detection is not necessary
for confirmation based on the BEL.) A high signal-to-noise composite
quasar spectrum from the FIRST Bright Quasar Survey \citep{Brotherton01} was used for line
identification. In most cases, an estimate of the quasar redshift can be
determined from a visual fitting of the emission features in evidence. Comparing emission line centers to line identifications from \citet{Brotherton01}, redshifts were determined to an accuracy of $\sigma_z \simeq several \times 10^{-3}$ in all cases. By far the main contribution to redshift uncertainty was the difference in calculated redshifts from multiple emission lines in the same quasar. Care must be taken when considering the redshift of a quasar determined from only one emission line. MgII is often the only emission line in evidence. There is a slight, but noticeable blueshift of the MgII line as a function of the quasar color \citep[e.g.][]{Richards03}.

\begin{figure}
\plotone{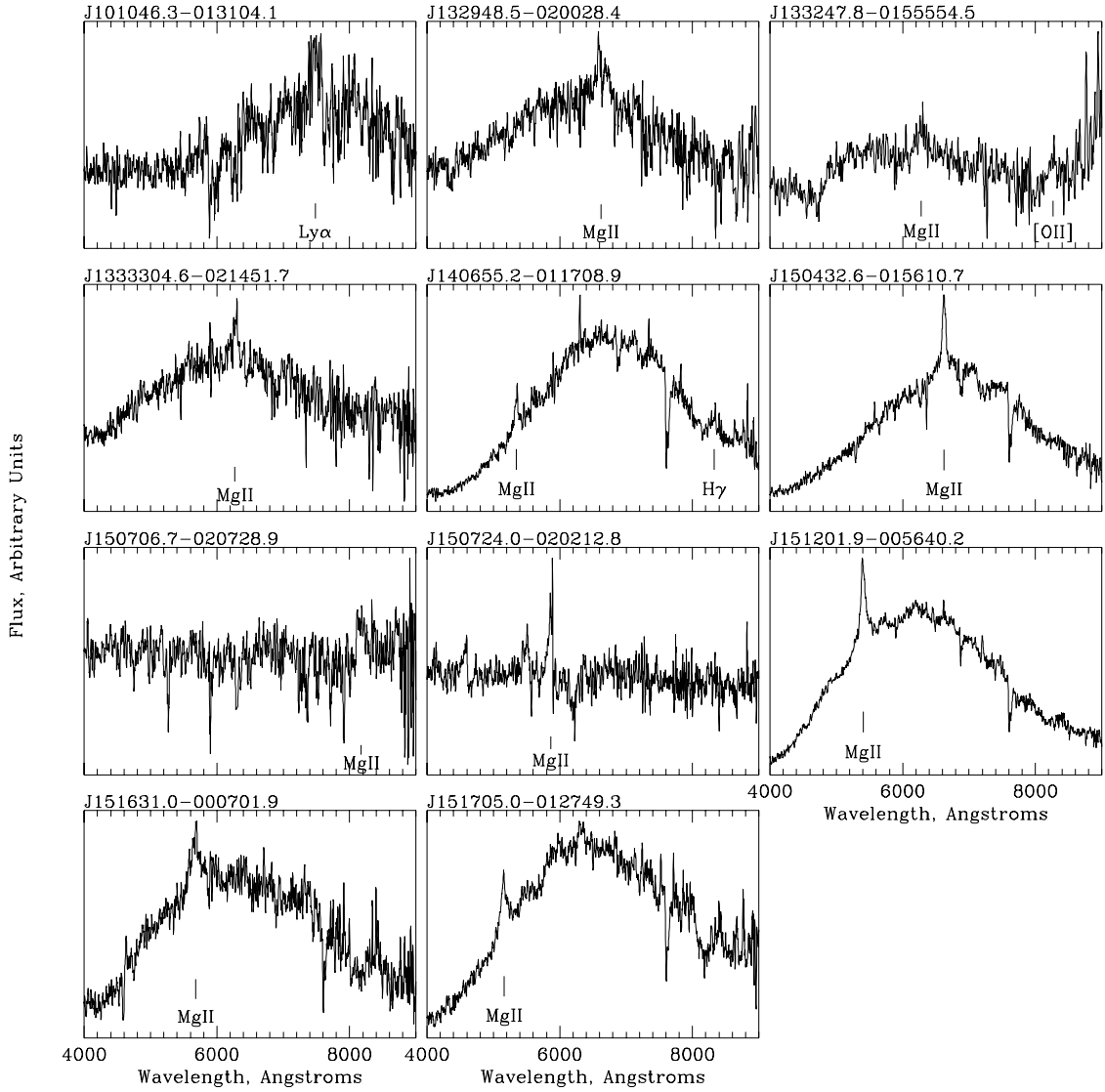}
\caption[Spectra of Newly Discovered Variable Quasars]{Low resolution Hydra spectra of all QUEST variable quasars not previously identified. The emission lines used for redshift determinations are labeled.}
\label{allspectra}
\end{figure}

Table\,\ref{qvq} lists the QUEST variable quasars discovered (or re-discovered) during the 2001 and 2002 spectral observing campaigns at the WIYN telescope. A total of 30 quasars were discovered during the two spectroscopic observing campaigns. During a subsequent literature search, 19 of these were found to be previously identified quasars and 11 are new discoveries. The spectra of each of the newly discovered variable quasars are given in Fig.\,\ref{allspectra}. The cross reference information given for previously discovered quasars is obtained with an object-by-object catalog search on the confirmed quasars using the NASA Extragalactic Database (NED). If a NED quasar is within $1\farcs0$ of a QUEST variable quasar, it is considered a rediscovery and the reference to the earliest known publication of that quasar is given. Also note that in two instances, a QUEST variable quasar has been matched to a published SDSS object for which there is published photometry, but not a spectrum \citep{Abazajian03}. Those are noted parenthetically with the SDSS photometric classification.

While reserving a full discussion of the selection effects of a variability-selected sample of quasars in relation to those of a UVX- or color-selected sample for a subsequent paper, it is worth commenting briefly on the color information for the subset of quasars in table\,\ref{qvq} that also have published SDSS photometry. 15 of the 30 variable quasars have published SDSS photometry and reliable color data \citep{Abazajian03}. Of those, 13 have been marked as quasar targets based on their colors. Of the remaining two, one was initially marked as being explicitly excluded based on its color (QUEST J133647.3-004856.5) but was subsequently considered as a SDSS serendipitous target and spectroscopically verified as a SDSS quasar and is also listed as an optically-selected LBQS quasar \citep{Hewett95}. The other quasar (QUEST J151631.0-000701.9) was classified as a photometric extended object by SDSS, but not flagged as a quasar candidate based on its colors. While 15 objects are not sufficient to make any substantive claims on the overlap or inherent selection effects in either selection process, it does begin to hint at the fact that the two populations of quasars are similar, but not identical. Further work and a much larger sample of quasars seen in both surveys will be necessary to make any quantitative claims as to the inherent biases and selection effects (if any) in the selection processes.

\subsection{Known QSO Variability}

In the 9th edition of the \citet{VC2000}, hereafter abbreviated as VC2000, there are over 13,000
known and cataloged quasars. Prior to direct spectroscopic
confirmation of the quasar candidates at WIYN, it was possible to test
our selection criteria by comparing the VC2000 catalog to the
variability-selected candidate list.

VC2000 is a compilation from over 40 sources of published
quasar and active galactic nuclei findings. The 9th edition contains
13,214 quasars, 462 BL Lac objects and 4,428 active galaxies and is
updated periodically. \citet{VC2000} define a quasar as an object
that is starlike or has a starlike nucleus, has broad emission lines
and has an absolute magnitude brighter than $M_{B} = -23$
\citep{Schmidt83}. VC2000 lists J2000 astrometry for each object, as
well as redshift, UBV or photographic photometry and radio (6 and 11
cm) flux densities when available.

Using only the data from the 1999 and 2000 epochs, there are 47 scans covering a time
baseline of 13 months. There are 159 VC2000 quasars within the scan
region bright enough to be detected in the QUEST scans and which have
a reliable astrometric match with an object in the candidate list. Of those
159 quasars, 141 have sufficient photometry to warrant further consideration
(i.e., objects that are seen through \emph{at least} both R filters).

In a previous study, \citet{Cimatti93} indicate that after one year,
$29\pm8\%$ of a sample of previously known quasars exhibited
variability. Following these results, with a time baseline of 13
months and a sample of 159 VC2000 known quasars, it would be expected that $46\pm13$ of those quasars would exhibit variability. Of the 141
VC2000 quasars under consideration, 53 (38\%) meet the variability
criteria explained above at the 85\% CL with an average redshift of $1.10\pm0.75$. Of those, 38 (27\%) also vary at the 95\% CL. These numbers are consistent with the findings of
\citet{Cimatti93} for a one-year baseline.

With the inclusion of the 2001 epoch data,
there are now 69 scans covering a total of 26 months. The structure
functions in \citet{Cristiani96} and \citet{Trevese94} indicate that quasar
variability seems to increase steadily for at least several years in
the quasar rest frame. With the caveat of redshift effects on the rest frame time baseline, doubling the time baseline from one year to two years should increase the number of quasars seen to be
variable. \citet{Cimatti93} report that $50\pm3\%$ of their quasars
were seen to vary over a time baseline of two years. 

Using the same selection and matching criteria as with the 13-month baseline, there are 153 VC2000 quasars which are matched to a QUEST candidate in the 26-month data. Of those, 137 are seen on
both R chips. As with the total number of detected objects, the difference in the number of detected VC2000 quasars (153 at 26 months versus 159 at 13 months) is a result of a cut in the data which requires an object to appear on a minimum percentage of the valid exposures. There are 83 ($61\%$) VC2000 quasars that meet the
variability criteria at the 85\% CL with an average redshift of $1.25\pm0.68$, 74 ($54\%$) of which also vary at
the 95\% CL. This is consistent with the work done by
\citet{Cimatti93}. It should be noted that, compared to the 13-month
data, the incremental increase in the 26-month data between the 85\% and
95\% CL is much smaller. That is to say, after 2 years, the vast
majority of the variable quasars that were detected are seen to vary
at a very high confidence level.

\section{Discussion}

\begin{figure}
\plotone{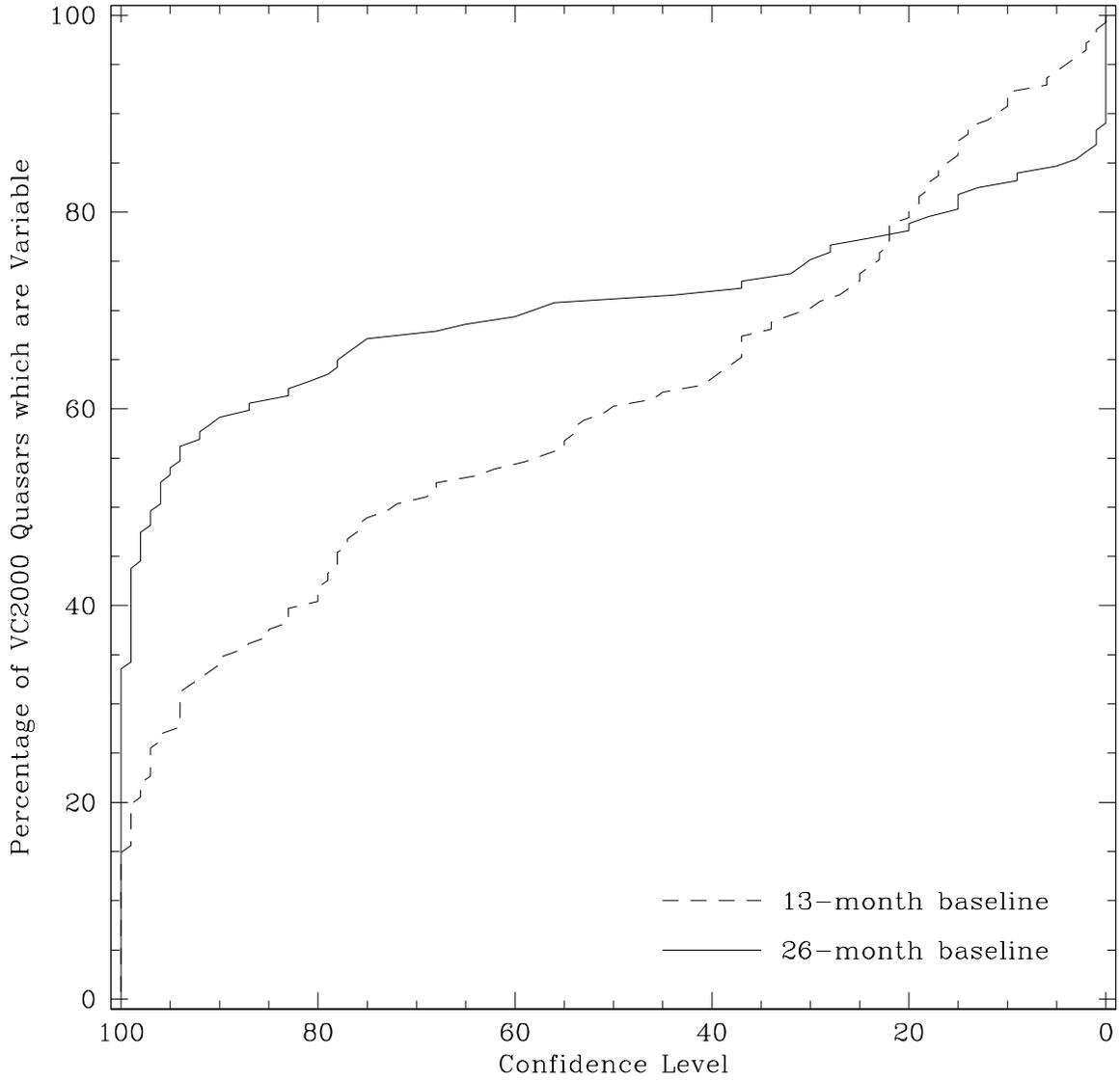}
\caption[Candidate-List Inclusion of VC2000 Quasars]{The percentage of VC2000 quasars which vary
  as a function of the minimum acceptable CL. The 13-month data are shown by the dashed curve and the 26-month data by the solid curve.}
\label{qsopercent}
\end{figure}

For the purposes of observing strategies and due to the limited telescope
time for spectroscopic follow-up attempts, the variability
candidate list was limited to those objects which have at least an 85\% CL. However, if the inclusion of VC2000 quasars
into the candidate list as a function of CL is analyzed \emph{a posteriori}, trends become visible as
both the CL and the time baseline change. Fig.\,\ref{qsopercent} shows
the completeness (i.e., the percentage of VC2000 QSOs determined to be variable) as a function of the CL. Both the 13-month and 26-month datasets are shown.

Two general trends are worth noting. Down to a relatively low
($CL\simeq20$) confidence level, the 26-month data show a higher completeness than the 13-month data. This is expected, under the assumption that the likelihood of variability
increases with time baseline. While both curves
monotonically increase, the 13-month data do not rise as rapidly at
high CL and they do not level off through the intermediate CL values,
as the 26-month data tend to. In the 13-month dataset, there is no
clear distinction between variable and non-variable quasars. A
gradually increasing number are accepted as variable as the CL
criterion is loosened. Contrast this with the 26-month dataset, in
which there is a high completeness even at a high CL
(over 50\% at the 95\% CL). With the increased time baseline, the
VC2000 quasars are beginning to show a marked distinction between
variability and non-variability. Decreasing the CL cutoff through its
intermediate values will not help in obtaining additional VC2000
quasars to the same extent that it does with the 13-month data
set. This conclusion supports our decision to limit the list of candidates for spectroscopic confirmation to a relatively high CL.

The second-generation QUEST camera has recently been installed on the
48-inch Oschin Schmidt telescope at Palomar Observatory and is
presently collecting data on both drift scan and `point and stare' mode.
The new site and telescope and the new camera design (112 thinned CCD
chips, $13 \mu$ pixels, $0.87$ arcsec per pixel) will all work to
improve the photometry to be obtained by QUEST in upcoming data
collection.  With the addition of a series of variability scans in the spring of
2004, the time baseline will increase from 26 months to $\sim60$
months. With a roughly $250\%$ increase in baseline and improved
photometry, a variability-selected sample will result in a more complete and more efficient list of objects suitable for spectroscopic confirmation as quasars. It is expected that the separation into a group of `definitely variable' QSOs and a group of `definitely nonvariable' QSOs, illustrated in Fig.\,\ref{qsopercent}, will also grow more distinct. In addition, The SDSS First Data Release \citep{Abazajian03} and the 2dF QSO Redshift Survey \citep{Croom03} will provide a more complete sample of published QSO spectra to which to compare the QUEST variable object list, resulting in more robust statistics concerning the completeness and efficiency of the selection criteria used in this and future work.

Future work in this field may also benefit from the study of the
light curve itself, either via statistical testing for the time frames
that contribute to variability \citep[eg.][]{Simonetti85} or higher-mode testing for intermittency in a nonlinear signal \citep[eg.][]{Vio92}. The addition of a zero proper motion \citep[eg.][]{Meusinger02} or color-based criteria \citep[eg.][]{Richards02} to the variability testing are also promising methods
of increasing the survey efficiency over variability alone, although understanding the implicit selection effects due to heterogeneous selection techniques may become an issue.

\section{Conclusions}

The few subjects touched on in the previous section show that
this project lends itself well to future directions for research. Additional
epochs of observational data can only improve the search for quasars
via variability. Various tests can illuminate the properties of the
variable objects already in hand. However, as a cohesive and complete
body of work, this research has proved fruitful and offers its own
conclusions.

A three-year campaign of high galactic latitude drift scan
observations was carried out at the Venezuelan National Observatory
from February 1999 to April 2001, resulting in 69 usable
observations. These data were processed and reduced using standard
data reduction algorithms, and analyzed for the presence of variable objects. 

The total number of ensemble objects dropped from over 300,000 after
13 months to about 280,000 after 26 months. The 24 scans taken during
2001 March and April effectively doubled the time baseline and 
increased the total number of scans by approximately 50\%. The
additional scans allowed for better statistics in both ensemble object
selection and variability determination and also allowed for more
reliable rejection of false positives. In addition, the increased
time baseline allowed for a greater fraction of known quasars to be
detected as variable objects.

The spectroscopic confirmation of a subset of
these variable objects showed that optical variability can be used as
a means of discovering, or re-discovering, quasars. During two
observing campaigns, a total of 30 quasars were detected, 9 of which
are new discoveries.

The thirty quasars discovered, or rediscovered, by this survey show
variability-selection of quasar candidates to be a qualified success. Both spectroscopic confirmation
observing campaigns proved to be more efficient than the statistics of
the variable VC2000 QSOs predicted. With a time baseline of only 26
months to date, variability as a means of quasar detection is expected to
increase in effectiveness as additional epochs of data are added to
the QUEST data. In the meantime, various methods of statstical
analysis can be investigated as a means to further improve the
efficiency of the spectroscopic observations.

\acknowledgments

This work is based on a Ph.D. dissertation submitted to Indiana University in August 2003 \citep{Rengstorf03} and was supported in part by a grant by the National Science Foundation to Indiana University. Adam Rengstorf would also like to thank Robert Brunner for insightful discussions about synoptic sky surveys.

\clearpage

\clearpage

\clearpage

\clearpage

\clearpage

\begin{deluxetable}{cccccccccc}
\tabletypesize{\footnotesize}
\tablewidth{444.19456pt}
\tablecaption{Summary of WIYN Spectral Observations\label{fields}}
\tablehead{ & & \multicolumn{2}{c}{Field Center} & & &
  \multicolumn{4}{c}{Type of Object \tablenotemark{a}} \\ \cline{3-4}
  \cline{7-10} \\
\colhead{Field} & \colhead{Date} & \colhead{RA} & \colhead{Dec} &
\colhead{Int. Time} & \colhead{\# of Spectra} & \colhead{Q}
& \colhead{S} & \colhead{M} & \colhead{?} \\ &
  & \colhead{(h m)} & \colhead{(\degr\ \arcmin)} & \colhead{Minutes}} 
\startdata
44 & 2001 Apr 18 & 10 31 & -1 21 & 60 & 16 & 2 & 11 & 1 & 2 \\
02 & 2001 Apr 18 & 12 36 & -0 21 & 60 & 42 & 1 & 26 & 10 & 5\\
01 & 2001 Apr 18 & 03 36 & -0 33 & 60 & 42 & 3 & 23 & 7 & 9 \\
04 & 2001 Apr 18 & 15 05 & -1 51 & 60 & 25 & 3 & 4 & 2 & 16 \\
03 & 2001 Apr 18 & 15 17 & -0 27 & 60 & 31 & 1 & 24 & 4 & 2 \\
31 & 2001 Apr 21 & 10 54 & -1 39 & 60 & 22 & 0 & 16 & 5 & 1 \\
08 & 2001 Apr 21 & 14 06 & -1 27 & 60 & 24 & 3 & 17 & 1 & 3 \\
33 & 2001 Apr 21 & 15 18 & -1 45 & 60 & 27 & 1 & 19 & 3 & 4 \\
10 & 2001 Apr 21 & 14 25 & -0 33 & 60 & 19 & 3 & 10 & 4 & 2 \\
11 & 2002 Feb 14 & 10 16 & -1 55 & 60 & 11 & 0 & 2 & 1 & 8 \\
16 & 2002 Feb 14 & 11 36 & -1 36 & 60 & 12 & 0 & 6 & 2 & 4 \\ 
10 & 2002 Feb 14 & 12 36 & -0 20 & 60 & 14 & 1 & 6 & 3 & 4 \\
13 & 2002 Feb 14 & 15 12 & -0 28 & 30 &  7 & 1 & 4 & 2 & 0 \\
39 & 2002 Feb 15 & 10 07 & -2 04 & 60 & 19 & 0 & 8 & 5 & 6 \\
09 & 2002 Feb 15 & 12 08 & -0 27 & 60 & 15 & 1 & 4 & 1 & 9 \\
02 & 2002 Feb 15 & 13 00 & -1 04 & 60 & 23 & 1 & 11 & 1 & 10\\
01 & 2002 Feb 15 & 13 36 & -1 02 & 60 & 24 & 2 & 16 & 3 & 3 \\
03 & 2002 Feb 15 & 15 17 & -0 25 & 60 & 27 & 0 & 25 & 1 & 0 \\
49 & 2002 Feb 16 & 10 13 & -1 42 & 60 & 12 & 1 & 5 & 5 & 1 \\
05 & 2002 Feb 16 & 13 31 & -2 05 & 74 & 22 & 4 & 12 & 4 & 2 \\
04 & 2002 Feb 16 & 13 44 & -0 21 & 60 & 18 & 4 & 11 & 3 & 0 \\
\enddata
\tablenotetext{a}{Q = Quasar; S = Stellar; M = M Dwarf; ? = Blank or Unknown}
\tablecomments{This table reports an internally prescribed field number, the UT date of the
observation and the J2000 coordinates of the field centers for each
observation, along with the total integration time, the number of
spectra obtained and a classification of each spectrum obtained into
one of four categories.}
\end{deluxetable}

\clearpage

\begin{deluxetable}{lccccccccl}
\tabletypesize{\scriptsize}
\tablewidth{0pt}
\tablecaption{QUEST Variable Quasar Catalog\label{qvq}}
\tablehead{\colhead{QUEST Identifier} & \colhead{RA} & \colhead{Dec} & \colhead{B} &
  \colhead{V} & \colhead{R} & \colhead{CL} & \colhead{z} &
  \colhead{z qual\tablenotemark{j}} & \colhead{Cross-reference} \\ &
  \colhead{(h m s)} & \colhead{(\degr\ \arcmin\ \arcsec)} & & & & & & & } 
\startdata
J101046.3-013104.1 & 10 10 46.3 & -1 31 04.1 & 19.89 & 18.13 & 17.33 & 100 & 5.09 & 0 & \nodata \\
J102956.9-012937.0 & 10 29 56.9 & -1 29 37.0 & 17.60 & 17.37 & 17.16 & 88  & 0.96 & 3 & LBQS 1027-0114\tablenotemark{a} \\
J103059.8-013237.7 & 10 30 59.8 & -1 32 37.7 & 19.19 & 19.18 & 18.84 & 97  & 2.15 & 5 & 2QZ J103059.7-013239\tablenotemark{b} \\
J120620.6-003851.7 & 12 06 20.6 & -0 38 51.7 & 18.74 & 18.71 & 18.66 & 100 & 0.56 & 4 & 2QZ J120620.5-003853\tablenotemark{b} \\
J123530.0-004140.9 & 12 35 53.0 & -0 41 40.9 & 19.04 & 18.77 & 18.44 & 95  & 1.60 & 4 & Q 1232-004\tablenotemark{c} \\
J130023.3-005429.2 & 13 00 23.3 & -0 54 29.2 & 18.68 & 18.00 & 17.66 & 100 & 0.12 & 5 & UM 534\tablenotemark{d} \\
J132948.5-020028.4 & 13 29 48.5 & -2 00 28.4 & 19.16 & \nodata & 18.29 & 96  & 1.37 & 0 & \nodata \\
J133141.0-015212.0 & 13 31 41.0 & -1 52 12.0 & 19.07 & 18.53 & 18.32 & 100 & 0.15 & 5 & 2dFGRS N198Z132\tablenotemark{e} \\
J133247.8-015554.5 & 13 32 47.8 & -1 55 54.5 & 19.14 & 19.07 & 18.73 & 97  & 1.23 & 1 & \nodata \\
J133304.6-021451.7 & 13 33 04.6 & -2 14 51.7 & 19.05 & \nodata & 18.50 & 99  & 1.24 & 0 & \nodata \\
J133635.3-002106.1 & 13 36 35.3 & -0 21 06.1 & 19.64 & 19.23 & 18.70 & 87  & 0.27 & 5 & SDSS J133635.26-002106.4\tablenotemark{f} \\
J133647.3-004856.5 & 13 36 47.3 & -0 48 56.5 & 17.57 & 17.50 & 17.28 & 100 & 2.80 & 4 & LBQS 1334-0033\tablenotemark{a} \\
J133649.2-002057.1 & 13 36 49.2 & -0 20 57.1 & 18.26 & 18.28 & 17.92 & 100 & 0.30 & 5 & LBQS 1334-0005\tablenotemark{a} \\
J133712.7-003927.7 & 13 37 12.7 & -0 39 27.7 & 18.91 & 18.58 & 18.38 & 100 & 1.51 & 2 & SDSS J133712.68-003928.2\tablenotemark{f} \\
J134255.7-004307.0 & 13 42 55.7 & -0 43 07.0 & 19.30 & 19.16 & 18.88 & 100 & 0.42 & 5 & 2QZ J134255.5-004308\tablenotemark{b} \\
J134256.6+000057.6 & 13 42 56.6 & +0 00 57.6 & 19.18 & 18.91 & 18.81 & 91  & 0.81 & 4 & SDSS J134256.51+000057.2\tablenotemark{f} \\
J134459.5-001559.0 & 13 44 59.5 & -0 15 59.0 & 17.93 & 17.87 & 17.54 & 100 & 0.25 & 4 & LBQS 1342-0000\tablenotemark{a} \\
J134548.0-002323.3 & 13 45 48.0 & -0 23 23.3 & 17.77 & 17.42 & 17.12 & 99  & 1.10 & 3 & LBQS 1343-0008\tablenotemark{a} \\
J140445.8-013021.6 & 14 04 45.8 & -1 30 21.6 & 18.69 & 18.39 & 18.14 & 99  & 2.51 & 4 & UM 632\tablenotemark{d} \\
J140621.6-012121.2 & 14 06 21.6 & -1 21 21.2 & 18.89 & 18.72 & 18.45 & 97  & 2.15 & 5 & 2QZ J140621.5-012122\tablenotemark{b} \\
J140655.2-011708.9 & 14 06 55.2 & -1 17 08.9 & 18.38 & 17.97 & 17.77 & 99  & 0.91 & 1 & \nodata \\
J142403.8-002657.5 & 14 24 03.8 & -0 26 57.5 & 16.60 & 16.29 & 16.07 & 100 & 0.15 & 5 & Q 1421-0013\tablenotemark{g} \\
J142600.2-002700.0 & 14 26 00.2 & -0 27 00.0 & 17.75 & 17.54 & 17.29 & 89  & 1.08 & 2 & EQS B1423-0013\tablenotemark{h} \\
J142658.6-002056.0 & 14 26 58.6 & -0 20 56.0 & 16.74 & 16.67 & 16.59 & 99  & 0.62 & 2 & EQS B1424-0007\tablenotemark{h} \\
J150432.6-015610.7 & 15 04 32.6 & -1 56 10.7 & 20.97 & 19.29 & 18.96 & 86  & 1.37 & 0 & \nodata \\
J150706.7-020728.9 & 15 07 06.7 & -2 07 28.9 & 18.24 & 17.31 & 16.77 & 99  & 1.92 & 0 & \nodata \\
J150724.0-020212.8 & 15 07 24.0 & -2 02 12.8 & 15.52 & 14.53 & 13.96 & 99  & 1.09 & 0 & \nodata \\
J151201.9-005640.2 & 15 12 01.9 & -0 56 40.2 & 18.62 & 17.90 & 17.65 & 99  & 0.93 & 0 & (SDSS GALAXY)\tablenotemark{i} \\
J151631.0-000701.9 & 15 16 31.0 & -0 07 01.9 & 19.67 & 19.66 & 19.26 & 86  & 1.03 & 0 & (SDSS STAR)\tablenotemark{i} \\
J151705.0-012749.3 & 15 17 05.0 & -1 27 49.3 & 19.72 & 19.33 & 19.10 & 99  & 0.84 & 0 & \nodata \\
\enddata

\tablenotetext{a}{LBQS Catalog \citep{Hewett95}}
\tablenotetext{b}{2dF QSO Redshift Survey \citep{Croom03}}
\tablenotetext{c}{Hewitt and Burbidge QSO Compilation
  \citep{Hewitt89}}
\tablenotetext{d}{Univ. of Michigan Emission Line Objects Catalog
  \citep{MacAlphine81}}
\tablenotetext{e}{2dF Galaxy Redshift Survey \citep{Baugh01}}
\tablenotetext{f}{SDSS Data Release 1 \citep{Abazajian03}}
\tablenotetext{g}{Veron-Cetty and Veron Catalog \citep{VC2000}}
\tablenotetext{h}{Edinburgh Quasar Survey Catalog \citep{Goldschmidt99}}
\tablenotetext{i}{SDSS GALAXY = extended object; SDSS STAR = point source; does not rule out future SDSS classification as a QSO.}
\tablenotetext{j}{redshift quality flag: 0 = single line ID w/out confirmation; 1 = multiple lines w/out confirmation; 2 = single line w/ confirmation $\Delta z < .03$; 3 = single line w/ confirmation $\Delta z < 0.01$; 4 = multiple lines w/ confirmation $\Delta z < 0.03$; 5 = multiple lines w/ confirmation $\Delta z < 0.01$}
\tablecomments{This table reports a standard catalog identifier, along with the right ascension and declination (J2000), B, V and R magnitudes, the variability confidence level, the calculated redshift and a flag denoting the quality of the redshift measurement.}
\end{deluxetable}

\clearpage

\end{document}